\documentclass[floatfix, twocolumn,amsmath,amssymb, aps,prb,superscriptaddress]{revtex4-2}
\usepackage{mathrsfs}
\usepackage{amsmath}
\usepackage{graphicx}
\usepackage{hyperref}
\usepackage{stfloats}
\usepackage{ulem}
\hypersetup{ colorlinks, citecolor=blue, linkcolor=blue, urlcolor=blue}

\begin{document}
	\title{Topological phase transition induced by band structure modulation in a Chern insulator}
	
	\author{Sayan Mondal}
	\affiliation{Department of Physics, Indian Institute of Technology Guwahati, 
		Guwahati 781039, Assam, India}
	\author{Priyadarshini Kapri}
	\author{Bashab Dey}
	\author{Tarun Kanti Ghosh}
	\affiliation{Department of Physics, Indian Institute of Technology Kanpur, 
	Kanpur 208016, U.P., India}
	\author{Saurabh Basu}
	\affiliation{Department of Physics, Indian Institute of Technology Guwahati, 
		Guwahati 781039, Assam, India}

\begin{abstract}
Here we study the systematic evolution of the topological properties of a Chern insulator in presence of an electronic dispersion that can be tuned smoothly from being Dirac-like till a semi-Dirac one and beyond. The band structure under such controlled deformation shows that the two Dirac points approach each other, merge at an intermediate point (the ${\mathbf{M}}$ point), where the low energy spectrum turns gapless, shows anisotropic Dirac features in the $k$-space and is denoted as the semi-Dirac limit, however a gap eventually opens up again in the spectrum. The Chern number phase diagram obtained via integrating the Berry curvature over the Brillouin zone (BZ) shows a gradual shrinking of the 'topological' lobes, and vanishes just beyond the semi-Dirac limit of the electronic dispersion. Thus there is a phase transition from a topological phase to a trivial phase across the semi-Dirac point. The vanishing
of the anomalous Hall conductivity plateau and the merger of the chiral edge states with the bulk bands near the ${\mathbf{M}}$  point provide robust support of the observed phase transition.
\end{abstract}
\maketitle
\section{Introduction}

In order to answer a question whether a crystal lattice with Bloch bands can possess a non-zero 
Chern number in the absence of an external magnetic field, Haldane in a seminal paper \cite{Haldane1988} had proposed a complex nearest neighbour 
hopping in a honeycomb lattice which explicitly breaks the time reversal symmetry (albeit no net flux) that yields a quantized anomalous Hall conductance, 
$\sigma_{xy}$. The quantization of $\sigma_{xy}$ occurs owing to the quantized (integral) values of the Chern number which results from the 
integral of the Berry curvature over the BZ. A non-vanishing Chern number and hence a finite Hall response are expected for a general value of the 
phase, $\phi$ of the next nearest neighbour hopping, where the closing (opening) of the energy gap occurs at the Dirac points ($\mathbf{K}$ and 
$\mathbf{K}^\prime$) depending on the value of the Semenoff mass, $m$. A rescaled variant of it, namely, $\tilde {m}$ ($= m/3\sqrt{3}\sin\phi$) 
as a function of $\phi$ shows a phase diagram that encodes opening and closing of the energy gaps alternately at the $\mathbf{K}$ and $\mathbf{K}^\prime$ 
points as $\tilde{m}$ is tuned between $+1$ and $-1$.

A striking distinction with the Dirac electrons (linear dispersion along both the longitudinal directions in $k$-space), is a case presented by an anisotropic dispersion where the bands are found to be linearly disperse along one direction and quadratically in the other \cite{Dietl2008,Banerjee2009}. Such inhomogeneities are found in TiO$_{2}$/VO$_{2}$ multilayered structures \cite{pickett2009, pickett2010}, where the band structures of the (TiO$_{2}$)$_{5}$/(VO$_{2}$)$_{n}$ ($n$ denotes the number of VO$_{2}$ layers) demonstrate the emergence of such a semi-Dirac dispersion for $n=3$ and $n=4$. In addition, semi-Dirac band dispersion is noted in a variety of other materials, such as monolayer phosphorene in presence of an electric field \cite{katnelson_2015, katnelson_2016}, doping and pressure \cite{castro_2014, guan_2014}, a deformed graphene structure \cite{montambaux_2009}, oxidized silicene layer \cite{zhang_2017} organic salts, such BEDT-TTF$_{2}$I$_{3}$ under pressure \cite{Suzumura2013,kohmoto_2006} etc. Further, several other features, such as the optical properties \cite{bryenton2019,mawrie2019}, Floquet states \cite{Narayan2015,Chen2018,Firoz2018}, thermoelectric properties\cite{Mandal2020,Mawrie2019(2)}, magnetic susceptibility, specific heat and Faraday rotation \cite{banerjee2012}, Landau levels and the transport properties etc \cite{Zhou2015,Yuan2016,sinha2020} are explored in literature for the semi-Dirac dispersion.

Motivated by the above scenario, here we have explored the evolution of the topological properties of a Haldane Chern insulator as one interpolates 
between the Dirac and the semi-Dirac dispersions. Such a possibility can be realized by choosing the hopping amplitude corresponding to one of 
the neighbours (say $t_{1}$) to be different than the other two (say $t$), where the former can be tuned starting from a value equal to the latter, that 
is, $t_{1}=t$ to larger values, for example, $t_{1}  \ge 2t$. In the regime $t < t_{1} < 2t$, we get a topological Chern insulating phase with a plateau 
at $e^{2}/h$ for the (anomalous) Hall conductivity, albeit the width of the plateau shrinks as it denotes the energy gap in the spectrum. 
A key feature of such an interpolation yields the pair of the Dirac points (the so called $\mathbf{K}$ and $\mathbf{K}^\prime$) 
in the electronic spectrum to move towards each other and eventually merge into a single band touching point, namely, the $\mathbf{M}$ point.  
At the $\mathbf{M}$ point the dispersion displays a linear behaviour in one direction and quadratic in the other when the value of $t_{1}$ equals $2t$. 
The $\mathbf{M}$ point is located at ($0,\frac{2\pi}{3a_{0}}$), $a_{0}$ being the lattice constant. Under this condition, the physics becomes similar to that 
of a single anisotropic Dirac cone in the low energy limit,
although the inversion symmetry is still preserved. This, along with a Berry phase of $\pi$ yields a finite Berry curvature at certain $k$-points in the BZ, 
despite not having any gap in the spectrum. 
The Chern number vanishes as soon as $t_{1}$ becomes greater than $2t$ where a trivial 
insulating phase emerges, thereby indicating a transition from a topological phase to a trivial one through a gap closing point. Appearance of two counter 
propagating modes at the edges is responsible for the emergence of a trivial insulating phase beyond the gap closing point.

Thus it remains to be seen, among other things, how the Chern number phase diagram and the anomalous Hall conductance respond to tuning of the band 
structure in a two dimensional honeycomb lattice from a Dirac to a semi-Dirac dispersion and even beyond that. In the following we perform a detailed  
study of the evolution of the band structure, density of states (DOS), edge modes in a nanoribbon, Berry curvature, Chern number phase diagram and finally the anomalous Hall conductivity to study the evolution of the topological properties. Importantly, we address the existence of a phase transition 
from a topological to a trivial phase embedded therein.

In section II we detail our model Hamiltonian that interpolates between a familiar topological Chern insulator introduced by Haldane to a trivial insulator as one of
the hopping amplitudes is progressively made larger compared to the other two. The evolution of the band structure and the density of states are plotted 
corresponding to these cases. Hence in section III we investigate the structure of the edge modes 
in a nanoribbon and specifically plot the energy spectrum in the vicinity of the 
$\mathbf{M}$ point to show their relevance in the ongoing discussion of the phase transition induced by engineering of the band structure. Hence we show evolution of 
the Berry curvature and the Chern number phase diagram in section IV which characterize the topological phase in a quantitative manner. Finally we show the anomalous
Hall conductivity in section V and conclude with a brief summary of our results in section VI.

\section{Model Hamiltonian}\label{sec_ham}

\begin{figure}[h]
	\centering
	\includegraphics[width=0.35\textwidth]{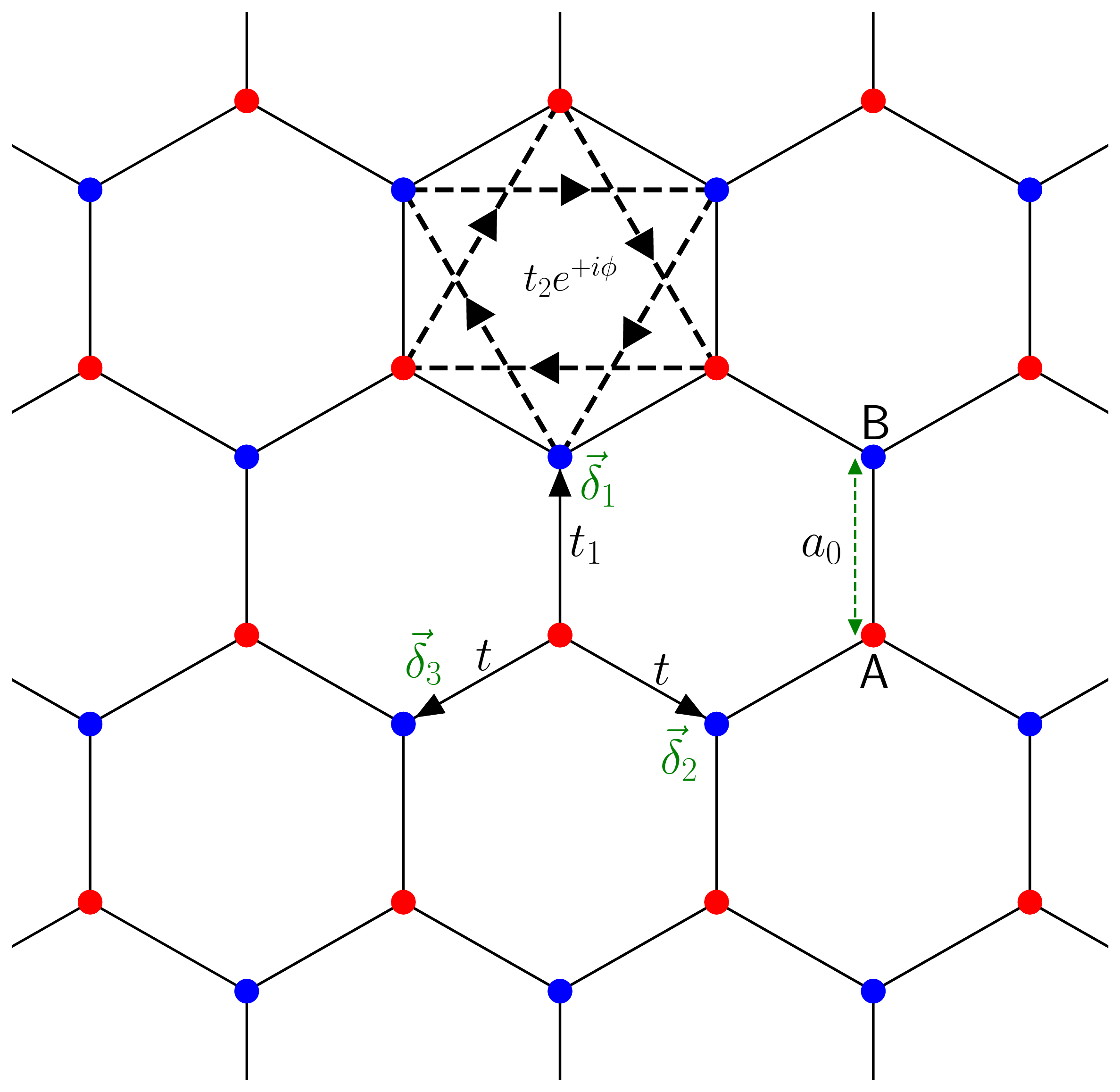}
	\caption{(Color online) A honeycomb lattice is shown where the red and the blue circles represent two different sublattices A and B respectively. 
		In the $\boldsymbol{\delta}_2$ and $\boldsymbol{\delta}_3$ direction, the nearest neighbour hopping strengths are same, namely, $t$,  whereas in the 
		$\boldsymbol{\delta}_1$ direction, it is $t_1$. At the top, the sign of the complex phase, $\phi$ corresponding to $nnn$ hopping is depicted.
		The arrows indicate the direction of the electrons hopping between the sites of the same sublattice (A to A or B to B). For the hopping to be 
		in the clockwise direction, $\phi$ is positive and for the hopping in the anti-clockwise direction $\phi$ is negative.}
	\label{lattice}
\end{figure}
We have considered a tight binding model on a honeycomb lattice as shown in Fig. \ref{lattice} with different nearest neighbour ($nn$) hopping strengths. The 
hopping strengths in two nearest neighbours directions ($\boldsymbol{\delta}_2$ and $\boldsymbol{\delta}_3$) is $t$, whereas in the third direction, that is, 
along $\boldsymbol{\delta}_1$, the strength is $t_1$. The $nn$ vectors are given by, $ \boldsymbol{\delta}_1 = a_0(0, 1)$, $\boldsymbol{\delta}_2= 
a_0(\sqrt{3}/2, -1/2)$ and $\boldsymbol{\delta}_3= a_0(-\sqrt{3}/2, -1/2)$.
In addition, we have introduced direction dependent next nearest neighbour ($nnn$) hopping,
$t_2e^{i\phi_{ij}}$ with an amplitude, $t_2$ and a complex phase $\phi$ having identical values along all the $nnn$ directions, where $\phi$ assumes
positive (negative) values if the electron hops in the clockwise (anti-clockwise) direction. 
The three $nnn$ vectors are denoted by $\boldsymbol{\nu}_1 = \boldsymbol{\delta}_2 - \boldsymbol{\delta}_3$, $\boldsymbol{\nu}_2 = 
\boldsymbol{\delta}_3 - \boldsymbol{\delta}_1$ and $\boldsymbol{\nu}_3 = \boldsymbol{\delta}_2 - \boldsymbol{\delta}_1$.
Following the Haldane model on a honeycomb lattice \cite{Haldane1988} the Hamiltonian for our system can be written as,
\begin{equation}\label{rawhamiltonian}
H = - \sum_{\left<i,j\right>} t_{ij} c_i^{\dagger} c_j + t_2 \sum_{\left<\left<i, j\right>\right>} e^{i\phi_{ij}} c_i^{\dagger} c_j + 
	\Delta_i\sum_i c_i^\dagger c_i + {\rm {h.c.}},
\end{equation}
where $c_i$ ($c_i^\dagger$) denotes the annihilation (creation) operators. The first, second and the third terms in Eq. (\ref{rawhamiltonian}) denote the $nn$
hopping, the complex $nnn$ hopping and the on-site energy terms, respectively. $\Delta_i$ is $+\Delta$ ($-\Delta$) depending on the A (B) sublattice sites. 
Performing a Fourier transform, the Hamiltonian in Eq. (\ref{rawhamiltonian}) can be written in the sublattice basis as,
\begin{eqnarray}\label{kphamiltonian}
H & = - &\left[ t_1 \cos(\mathbf{k}\cdot\boldsymbol{\delta}_1) + \sum_{i = 2}^3 t\hspace{3pt} \cos(\mathbf{k}\cdot\boldsymbol{\delta}_i)\right]\sigma_x   \\ \nonumber
& - & \left[t_1 \sin(\mathbf{k}\cdot\boldsymbol{\delta}_1) + \sum_{i = 2}^3 t\hspace{3pt} \sin(\mathbf{k}\cdot\boldsymbol{\delta}_i) \right] \sigma_y   \\ \nonumber
& + & \left[\Delta -2\hspace{3pt}t_2\hspace{3pt}\sin\hspace{2pt}\phi\sum_{i=1}^{3} \sin(\mathbf{k}\cdot\boldsymbol{\nu}_i) \right]\sigma_z \\ \nonumber
& +  & \left[2t_2 \hspace{3pt}\cos\hspace{2pt}\phi\sum_{i=1}^{3} \cos(\mathbf{k}\cdot\boldsymbol{\nu}_i) \hspace{3pt}\right]I \\ \nonumber 
& = & h_x \sigma_x + h_y \sigma_y + h_z \sigma_z + h_0 I ,
\end{eqnarray}
where $\sigma_{i}$ are the $2\times2$ spin-$1/2$ Pauli matrices which denote the sublattice degrees of freedom and $I$ is the $2\times2$ identity matrix.
$h_{i}$ as shown in square brackets in Eq. (\ref{kphamiltonian}) are the coefficients of $\sigma_{i}$.
The energy dispersion can be obtained as, $E(k)= h_0\pm[h_x^2 + h_y^2 + h_z^2]^{1/2}$, where the $\pm$ signs correspond to the conduction and the valence band
dispersions respectively. 
\begin{figure}[h]
\includegraphics[width = 0.48\textwidth]{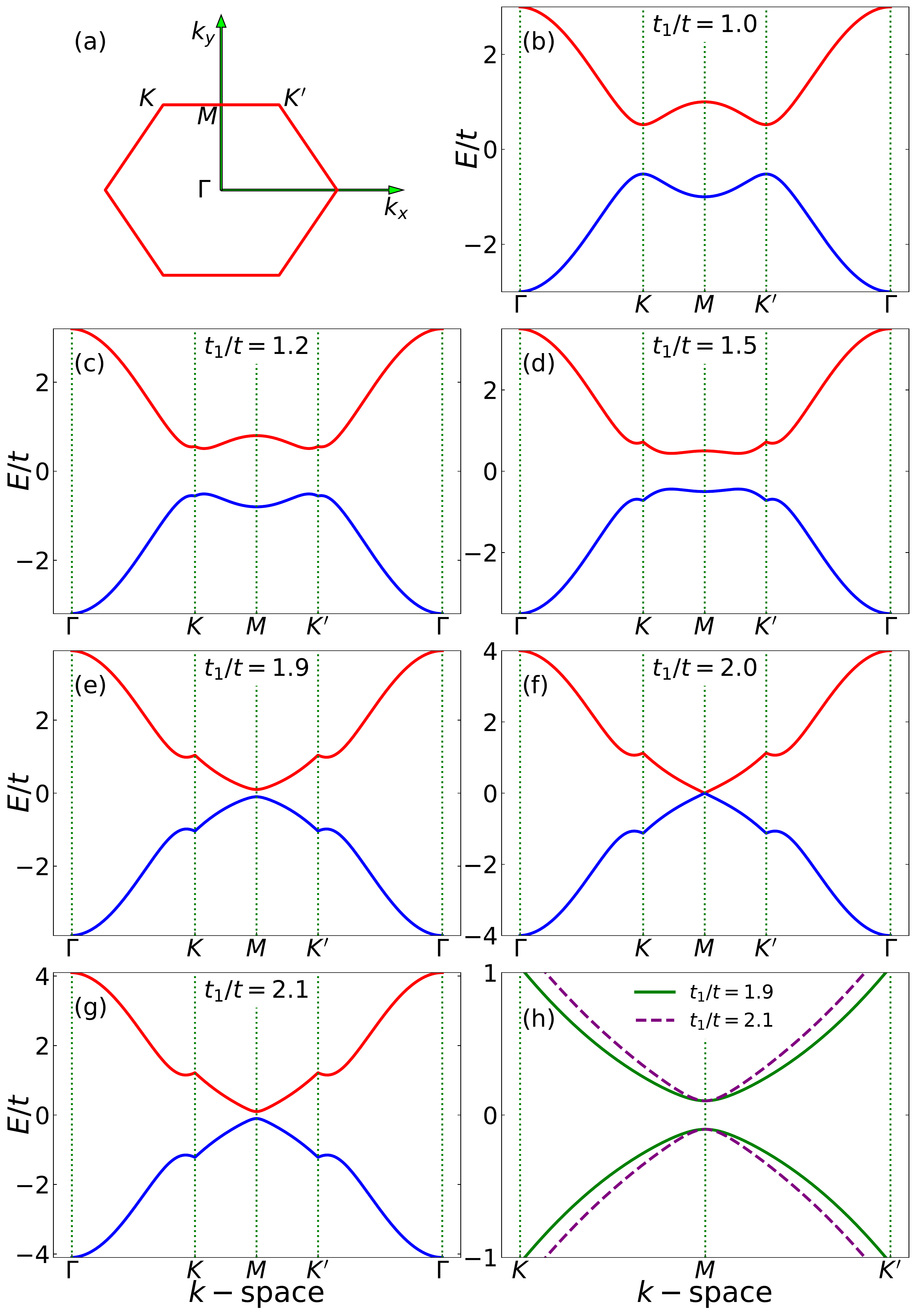}
\caption{(a) Brillouin zone of a honeycomb lattice, the electronic band dispersion in the $k$-space along the 
$\Gamma \rightarrow K \rightarrow M \rightarrow K^\prime 
\rightarrow\Gamma$ direction for (b) $t_1 = t$, (c) $t_1 = 1.2t$, (d) $t_1 = 1.5t$, (e) $t_1 = 1.9t$, (f) $t_1 = 2t$ and (g) $t_1 = 2.1t$ are shown. 
(h) A closer view of the gap for $t_1 = 1.9t$ and $t_1 = 2.1t$ along $K\rightarrow M\rightarrow K^\prime$ direction are presented. For all the figures 
$t_2$, $\Delta$ and $\phi$ are kept fixed at $0.1t$, $0$ and $\pi/2$ respectively.}
\label{bandstructure}
\end{figure}
To make the notations on different symmetry points clear, the BZ of the honeycomb lattice and the electronic band structures corresponding to different values of
$t_1$ are shown in Fig. \ref{bandstructure}. The $nnn$ hopping amplitudes and
the complex phase are fixed all the while at $t_2 = 0.1t$ and $\hspace{3pt}\phi = \pi/2$ respectively. Inclusion of this complex $nnn$ hopping opens up a gap, 
thereby making it an insulating state. For $t_1 = t$ and $t_2\neq 0$, the system is the famous Haldane model \cite{Haldane1988, Vanderbilt}, where a gap opens at 
the two non-equivalent Dirac points, that is, at the $\mathbf{K}^\prime$ $(2\pi/3\sqrt{3}a_0, 2\pi/3a_0)$ and at the $\mathbf{K}$ $(-2\pi/3\sqrt{3}a_0, 2\pi/3a_0)$ 
points (see Fig. \ref{bandstructure}b). As $t_1$ is tuned to larger values, these two points come closer to each other and along with a diminishing 
band gap occurs in the spectrum. As $t_1$ becomes equal to  $2t$, that is the semi-Dirac limit, the two bands touch at the $\mathbf{M}$ $(0, 2\pi/3a_0)$ 
point (see Fig. \ref{bandstructure}f), even in the presence of  $t_2$. This is obvious from the expression of $h_z$, which is zero only at the 
$\mathbf{M}$ point, becomes non-zero as one moves away from the $\mathbf{M}$ point and eventually at the $\mathbf{K}$ or the $\mathbf{K}^\prime$ 
points it reaches its maximum value, namely, $|3\sqrt{3} t_2|$.  When $t_1 = 2.1t$, again a gap opens up in the spectrum
as $h_x$ and $h_y$ now become non-zero at the $\mathbf{M}$ point. Moreover, the bandwidth increases from $6t$ in the Dirac case to 
$2(2t + t_1)$ for the semi-Dirac one.

\begin{figure}[h]
\centering
\includegraphics[width = 0.48\textwidth]{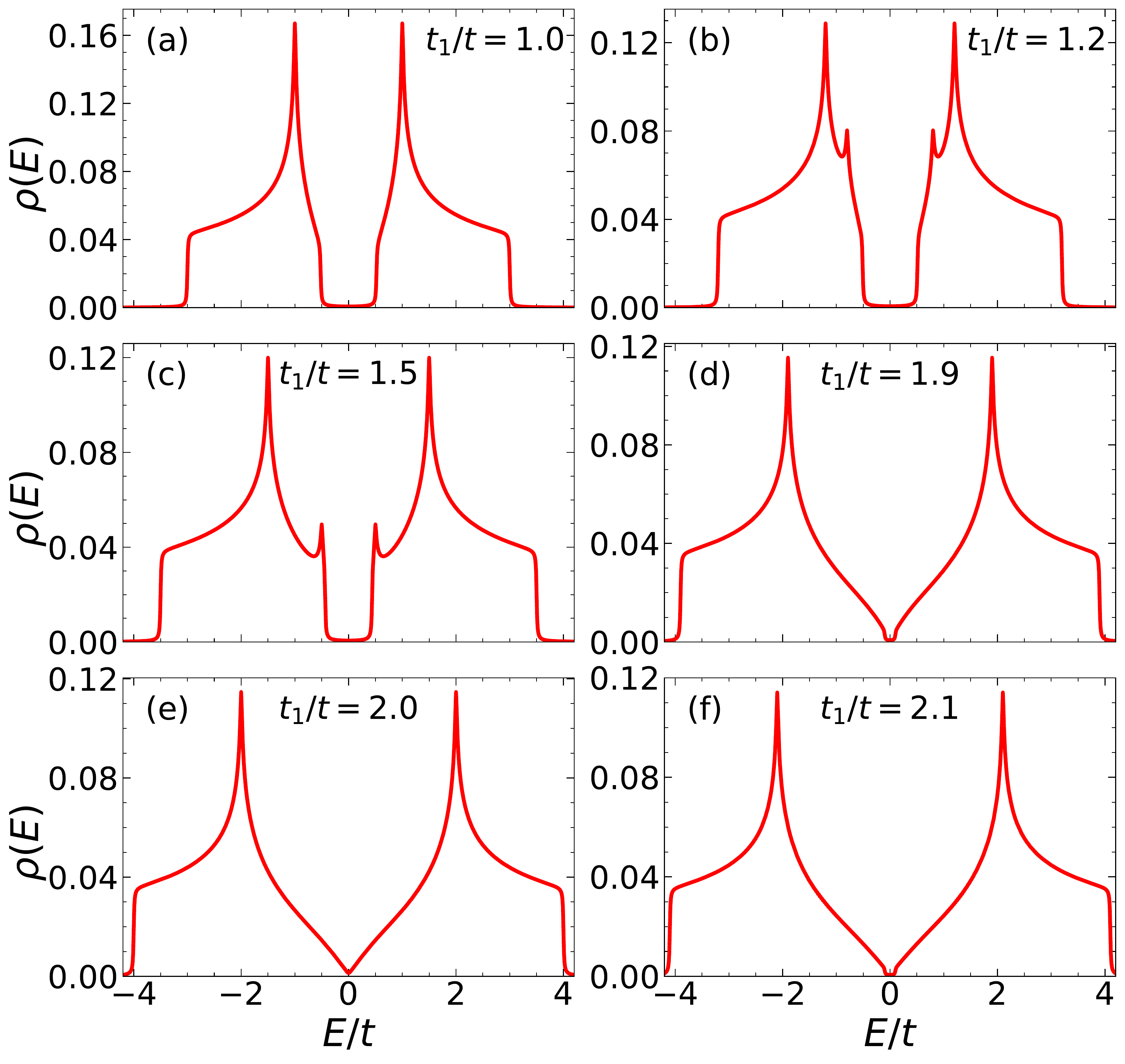}
\caption{The density of states (DOS) (in units of $1/t$) is plotted as a function of $E/t$ for (a) $t_1 = t$, (b) $t_1 = 1.2t$, (c) $t_1 = 1.5t$, 
(d) $t_1 = 1.9t$, (e) $t_1 = 2t$ and (f) $t_1 = 2.1t$. Through out the calculations $t_2$, $\Delta$ and $\phi$ are kept fixed at $0.1t$, $0$ and 
$\pi/2$ respectively.}
\label{DOS}	
\end{figure}

Next we compute the density of states (DOS) that counts the number of electronic states in the vicinity of a particular value of energy $E$. 
It can be obtained from the following relation,
\begin{equation}
\rho(E) = \frac{1}{(2\pi)^2}\int_{BZ} \delta(E - E(\mathbf{k})) d\mathbf{k}.
\end{equation}
Hence we have numerically computed the DOS for several different values of $t_1$ as shown in Fig. \ref{DOS}.
Two extra peaks (other than the dominant ones whose origin is well understood) in the DOS spectrum appear for $t_1 = 1.2t$ and $t_1 = 1.5t$ 
(see Fig. \ref{DOS}b and \ref{DOS}c) because of the presence of saddle points 
in the band structure (see Fig. \ref{bandstructure}c and \ref{bandstructure}d) occurring at the $\mathbf{M}$ $(0, 2\pi/3)$ point in the BZ. The saddle points are
absent for larger values of $t_{1}$, namely, $t_1 \ge 1.9t$ which causes the the additional peaks to be absent in the DOS spectrum. These peaks would occur 
even in absence of Haldane flux \cite{sinha2020}. For $t < t_{1} < 2t$, the DOS becomes non-vanishing in the vicinity of the zero energy which 
is also evident from the band structure shown in Fig. \ref{bandstructure}. As soon as $t_1$ becomes equal to $2t$, the DOS vanishes quadratically 
near $E/t = 0$ (Fig. \ref{DOS}d). Beyond that (at $t_{1}=2.1t$), the DOS again becomes zero at $E\simeq 0$ and thus implies of a gap opening up in the spectrum.

\section{Edge States}
So far we have looked at the model of infinite honeycomb lattice which is periodic in both the $x$ and the $y$-directions. However edge states are
best perceived in a 
system is of finite size. It is therefore interesting to consider a semi-infinite ribbon with a finite number of sites in the $y$-direction and an infinite 
number of sites in the $x$ direction, such that we label the sublattices as A$_1$, B$_1$, A$_2$, B$_2$, .... A$_N$, B$_N$ etc along the $y$-direction. 
The periodicity and hence the translational symmetry of the ribbon in the $y$-direction are broken because of the edges being present, while along the
$x$-direction the symmetry is preserved.  So we can Fourier transform the operators only in the $x$-direction, namely, use 
$c_{x, y}^\dagger = \sum_{k} e^{ikx}c_{k, y}^\dagger$. Incorporating such transformation in the tight binding Hamiltonian (Eq. (\ref{rawhamiltonian})), 
we arrive at the following two eigenvalue equations,
\begin{equation}\label{edge1}
\begin{aligned}
E_{k} a_{k, n} =&-\left[ t\left\{1+ e^{(-1)^n ik} \right\}b_{k, n} + t_1 b_{k, n-1} \right]\\ &-2t_2\left[ \cos(k + \phi)a_{k, n} + e^{(-1)^n\frac{ik}{2}}\times \right.\\ &\left. \cos\left( \frac{k}{2} - \phi\right) \{a_{k, n-1} + a_{k, n+1} \} \right]
\end{aligned}
\end{equation}
\begin{equation}\label{edge2}
\begin{aligned}
E_{k} b_{k, n} =&-\left[ t\left\{ 1+ e^{(-1)^{n+1} ik} \right\}a_{k, n} + t_1 a_{k, n+1} \right]\\ &-2t_2\left[ \cos(k - \phi)b_{k, n} + e^{(-1)^{n+1} \frac{ik}{2}}\times \right.\\ & \left. \cos\left( \frac{k}{2} + \phi\right)\{a_{k, n-1} + a_{k, n+1}\} \right]
\end{aligned}
\end{equation}
\begin{center}
\begin{figure}[h]
\includegraphics[width = 0.48\textwidth]{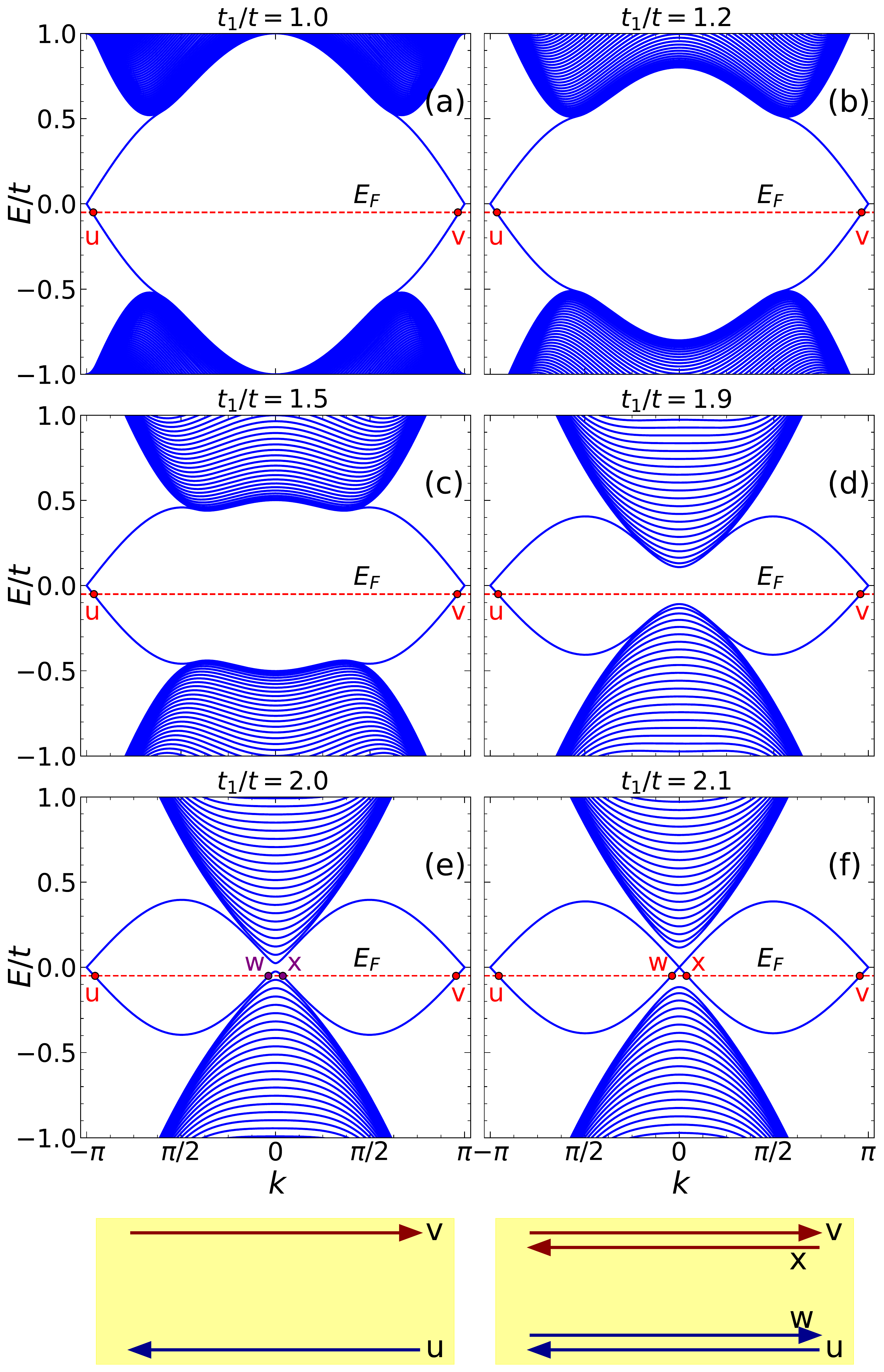}
\caption{The energy spectra  as a function $k$ (here $k$ denotes $k_{x}$) are shown for (a) $t_1 = t$, (b) $t_1 = 1.2t$, 
(c) $t_1 = 1.5t$, (d) $t_1 = 1.9t$, (e) $t_1 = 2t$ and 
(f) $t_1 = 2.1t$. In each figure, the edge state are shown by red dots at the Fermi energy $E_F$ (red dashed line), while in (e) the purple dots are the 
bulk states. A schematic diagram of a part of the ribbon with edge current are shown at the bottom of the (e) and (f) in yellow panels, where the 
edge currents along the edges are depicted by arrows. For $t_1 = 2.1t$, along one edge there are two edge states, whereas for $t_1\leq2t$ there is 
only one edge state. Through out the calculations $N$, $t_2$, $\Delta$ and $\phi$ are kept fixed at $128$, $0.1t$, $0$ and $\pi/2$ respectively.}
\label{figedge}
\end{figure}
\end{center}
where $n$ stands for the $n$-th sublattice, that is, $n$ can be an integer in the range $[1:N]$. $N$ is related to the width of the ribbon 
(see below) and $k$ is the (dimensionless) momentum 
in the $x$-direction defined by $k = \sqrt{3}a_0 k_x$. In Eqs. (\ref{edge1}) and (\ref{edge2}), $a_{k, n}$ and $b_{k, n}$ are the coefficients of the 
wavefunctions corresponding to the sublattices A and B respectively. By solving these two equations, we have obtained the band structure of the ribbon
for different values of $t_1$ in the range $[t:2t]$ and also for $t_{1} > 2t$ as shown in Fig. \ref{figedge} for a given value of $N$. 
The width of the ribbon can be obtained calculated from the relation 
$W(N) = a_0\left(\frac{3N}{2} -1\right)$. In our work, we have used $N=128$ and hence the ribbon has a width of 191$a_0$ in the $y$-direction. 
As can be seen, one of the modes from the lower band crosses over to the upper band with increasing values of $k_x$ and another one crosses in the opposite 
direction. The amplitudes of the wavefunction of these modes decay exponentially into the bulk from a maximum value at the edge of the ribbon 
\cite{nakada1996,castroneto,basu2018}. It should be noted that the velocities have opposite signs (since $v$ is proportional to $\partial E/\partial k$) along these states, 
implying that the electrons move in opposite directions along the edges, which makes the edge states chiral. Owing to the presence of the edge states 
there will be a plateau in the Hall response of value $e^2/h$. The edge states are shown for various values of $t_1$ in Fig. \ref{figedge}.

In Fig. \ref{figedge}e, we have shown the band structure for $t_1 = 2t$, where the intersection of the edge states with the Fermi energy, $E_F$ (shown via a
dashed line at $E/t = 0$) are represented by the dots \textit{u} and \textit{v}, whereas the dots \textit{w} and \textit{x} belong to the bulk states. Below the
figure, the edge currents corresponding to the points \textit{u} and \textit{v} along the edges of the ribbon are depicted in a yellow panel. 
Although there is a small gap discernible at $k=0$ for $t_{1}=2t$, if we gradually increase the value of $N$, or equivalently, the width of the ribbon, we shall
observe vanishing of the gap. Consequently in an infinite system there will be no gap in the band structure as inferred from Fig. \ref{bandstructure}f.
A similar picture of the edge states for $t_1 > 2t$, that is at $t_{1} = 2.1t$ is depicted in Fig. \ref{figedge}f, where unlike the previous one, 
the points \textit{w} and \textit{x} denote the two edge states. Therefore, along each edge, there are two counter-propagating modes as 
shown at the bottom of Fig. \ref{figedge}f.

Such pairs of oppositely moving helical edge modes (in Fig. \ref{figedge}e) are familiar in the context of quantum spin Hall (QSH) phase where each mode at 
either edge correspond to one or the other kind of spin. Since we have not included the spin degrees of freedom, there is no relevance of the QSH phase. However it 
accounts for vanishing of the anomalous Hall response due to cancelation of the Hall current. Incidentally, pair of such counter propagating 
chiral modes are observed in the trivial insulating phase in the Haldane model of a dice lattice \cite{Kapri2020}.

\section{Chern Number}\label{sec_Chern}
\begin{figure}[h]
\includegraphics[width = 0.42\textwidth]{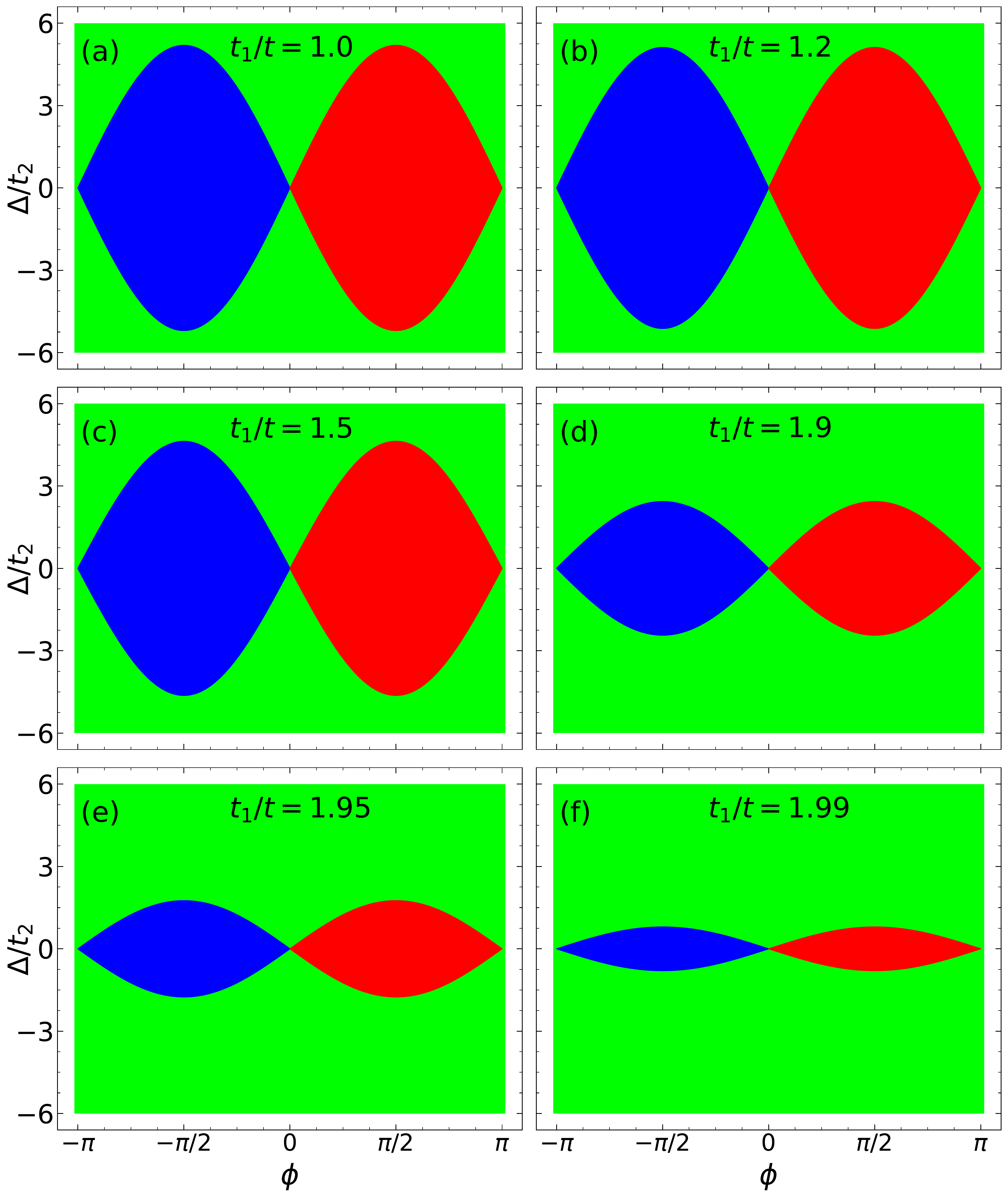}
\caption{Phase diagram for (a) $t_1/t = 1$, (b) $t_1/t = 1.2$, (c) $t_1/t = 1.5$, (d) $t_1/t = 1.9$, (e) $t_1 = 1.95t$ and (f) $t_1 = 1.99t$. 
Each of the red and the blue regions represent Chern insulating phase with Chern number +1 and -1 respectively, whereas the green region in each figure is the normal insulating phase with Chern number 0.}
\label{phasediag}
\centering
\end{figure}
\begin{figure}[h]
\includegraphics[width = 0.42\textwidth]{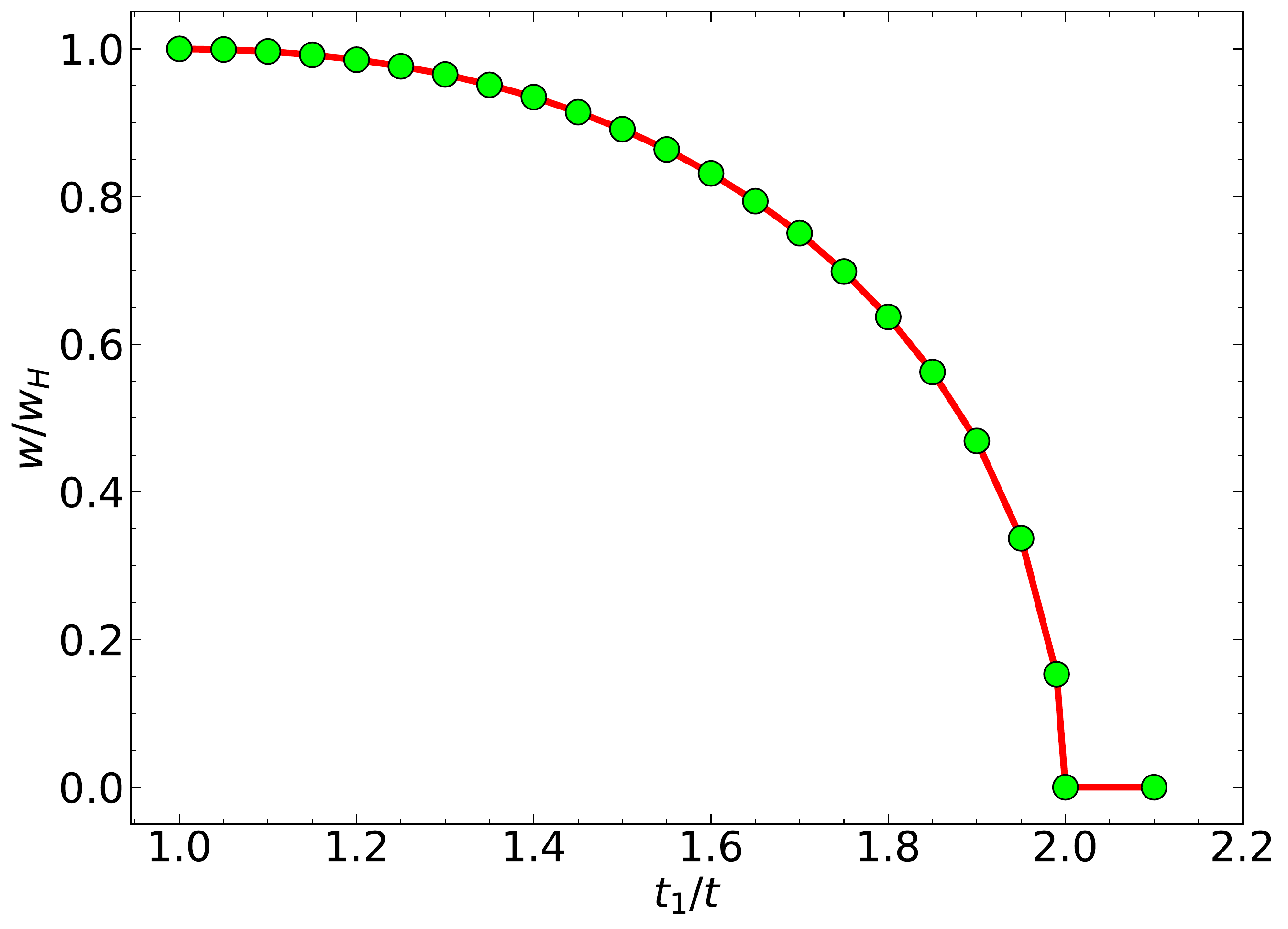}
\caption{The width of the Chern insulating phase as a function of $t_1/t$ corresponding to $\phi = \pi/2$ is shown. 
Here $w_H = 6\sqrt{3}t_2$ is the width for $t_1 = t$, that is, the width of the lobe in the Haldane model.}
	\label{cw}
\end{figure}
\begin{figure}[h]
\centering
\includegraphics[width=0.45\textwidth]{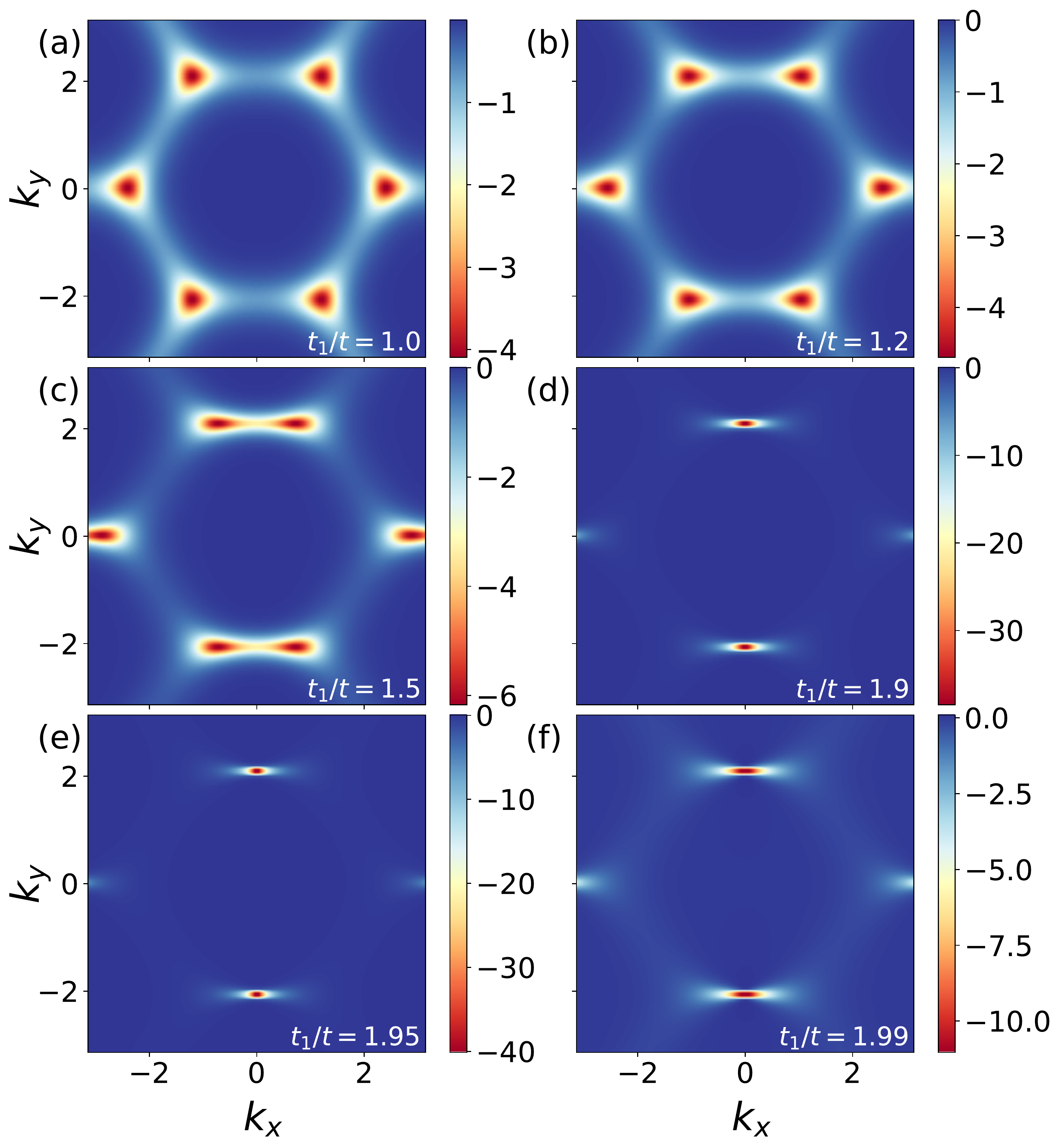}
\caption{Berry curvatures are shown in the $k_{x} - k_{y}$ plane for (a) $t_1 = t$, (b) $t_1 = 1.2t$, (c) $t_1 = 1.5t$, (d) $t_1 = 1.9t$, 
(e) $t_1 = 1.95t$ and (f) $t_1 = 1.99t$.
Here $t_2$, $\phi$ and $\Delta$ are fixed at $0.1t$, $\pi/2$ and $0$ respectively. The colorbar denotes the magnitude of the Berry curvature.}
\label{berrycurv}
\end{figure}

As we have already discussed in section \ref{sec_ham}, there are two types of hopping, $nn$ hopping amplitudes ($t_{1}$ and $t$)  and a complex 
$nnn$ hopping, $t_2 e^{\pm i\phi}$. The complex $nnn$ hopping term breaks the time reversal symmetry which is essential to obtain a non-zero Chern number. 
Now we turn on the inversion-symmetry breaking on-site energies $+\Delta$ and $-\Delta$ at the sublattices A and B respectively \cite{semenhoff}. 
Tuning this parameter we can open or close an energy gap.
\begin{figure}[b]
	\centering
	\includegraphics[width = 0.45\textwidth]{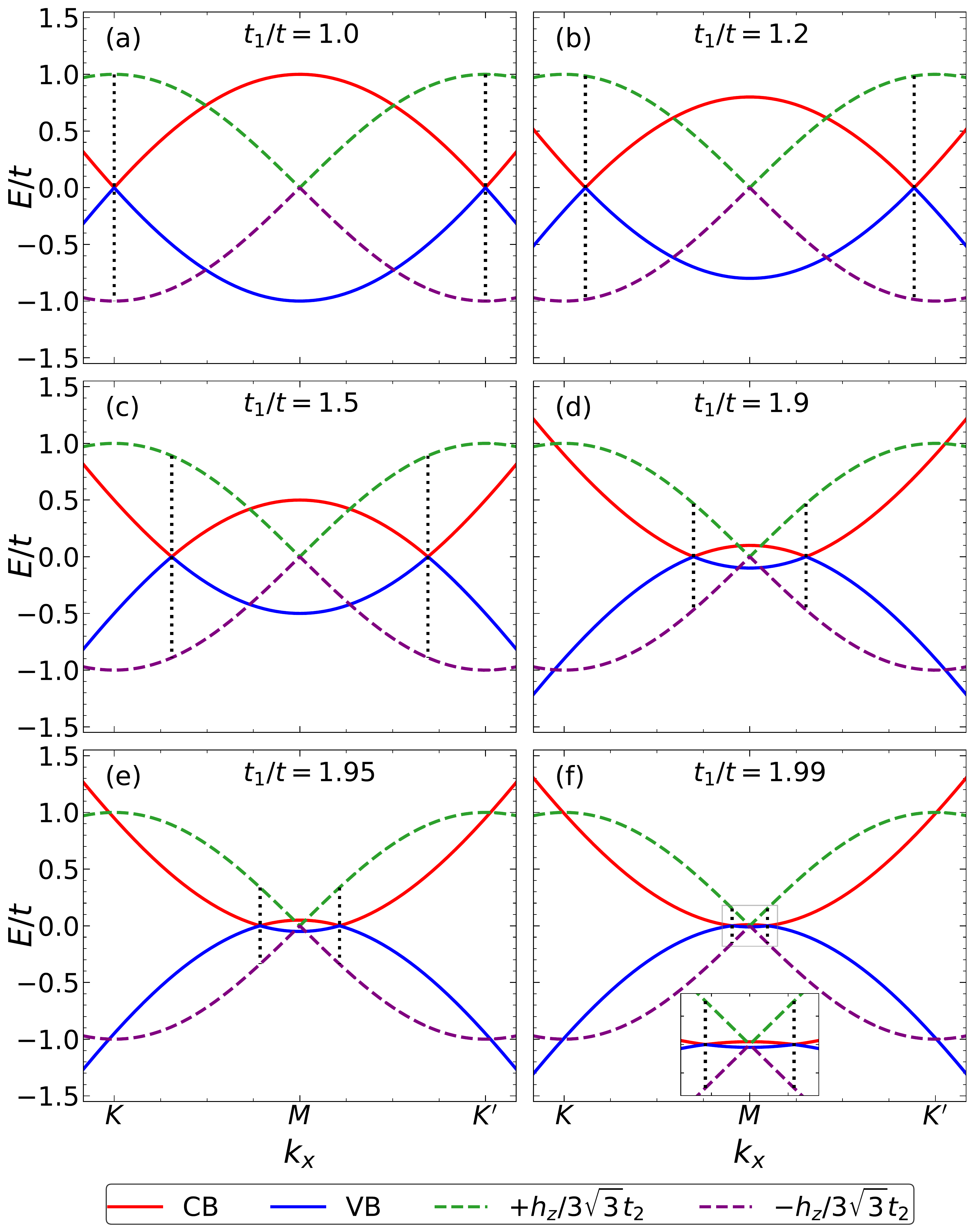}
\caption{The band dispersion in the absence of the $nnn$ hopping $(t_2 = 0)$ along $K\rightarrow M \rightarrow K^\prime$ direction, that is,
the $k_{x}$ direction for (a) $t_1 = t$, (b) $t_1 = 1.2t$, (c) $t_1 = 1.5t$, (d) $t_1 = 1.9t$, (e) $t_1 = 1.95t$ and (f) $t_1 = 1.99t$.
In the inset of (f), a closer view in the vicinity of $\mathbf{M}$ point is shown. In each of the figures, the red and the blue curves represent 
the conduction band (CB) and the valence band (VB) respectively, while the green and the purple dashed curves are positive and negative values of 
$h_z$ scaled by $3\sqrt{3}t_2$. The vertical dotted lines are plotted at $k_{x}$ points where the CB and VB touch each other. Here we have set $\phi$ and 
$\Delta$ as $\pi/2$ and $0$ respectively.}
\label{Chern_band}
\end{figure}
Now in order to get information about the non-trivial phase in such a scenario, we have used the Hamiltonian in Eq. (\ref{kphamiltonian}) and calculated the
Chern number numerically by the method discussed in Ref. \cite{Fukui}. A brief outline of the method appears in the following.
One can compute the Berry curvature in the $k_{x}-k_{y}$ plane which is is defined by,
\begin{eqnarray}
{\bf {\Omega}}({\bf {k}}) & = & \nabla_{{\bf {k}}}\times\boldsymbol{\cal {A}}({\bf {k}}) \\ \nonumber
& = & i\left[\left< \frac{\partial \psi({\bf {k}})}{\partial k_{x}}\right|\left. \frac{\partial \psi({\bf {k}})}{\partial k_{y}}\right> - 
\left<\frac{\partial \psi({\bf {k}})}{\partial k_{y}}\right|\left. \frac{\partial \psi({\bf {k}})}{\partial k_{x}}\right> \right] 
\end{eqnarray}
The $t_{1} = t$ case (Fig. \ref{berrycurv}a) represents a familiar scenario where the non-zero (and negative) values are highly concentrated around the 
Dirac points which persists as $t_{1}$ starts deviating from $t$ till moderate values of $t_{1}$.

The Chern number can now be calculated by integrating the Berry curvature over the BZ using \cite{thouless}, 
\begin{eqnarray}
	C & = & \frac{1}{2\pi}\int\int_{BZ}{\mathbf {\Omega}}({\bf {k}})\cdot{\bf {dS}} 
\end{eqnarray}
For a given value of $t_1$, we have obtained the Chern number corresponding to the lower  
band as a function  of $\Delta$ and $\phi$, keeping $t_2$ constant as shown in the phase diagrams of Fig. \ref{phasediag}. As can be seen, 
there are three regions in each phase diagram. The red ($C = +1$) and the blue ($C = -1$) regions are the topologically non-trivial insulating phase, 
while the green region ($C = 0$) denotes a trivial insulator. When $t_1 = t$ we get the phase diagram of the Haldane model \cite{Haldane1988, Vanderbilt}
(see Fig. \ref{phasediag}a), where the phase transition occurs at $\Delta = |3\sqrt{3}t_2|$ for $\phi = \pi/2$. Whereas if we look at the phase diagrams 
(Fig. \ref{phasediag}b - \ref{phasediag}e) for other values of $t_1$, we obtain what looks similar to that of the Haldane model except for the size of Chern 
insulating region gradually shrinks. These results are consistent with the band structure plots (see Fig. \ref{bandstructure}). As the band gap 
decreases with the increasing strength of $t_1$, phase transitions between the topological and the trivial phases occur for lesser values of $\Delta$. 
In the semi-Dirac limit, the energy spectrum consists of gapless anisotropic Dirac cone in the low energy limit.
Here $t_{1} = 2t$ denotes a critical point (gap closing point) which demarcates the topological phase ($t < t_{1} < 2t$) from a normal insulating phase
($t_{1} > 2t$). As the value of $t_{1}$ becomes larger than $2t$, that is, $t_1 = 2.1t$, the system becomes a normal insulator.

The shrinking of the width of the Chern lobes in shown in Fig. \ref{cw}. 
As can be seen from Fig. \ref{phasediag}, the width is maximum at $t_1 = t$ and has a value $6\sqrt{3}t_2$ (denote it by $w_{H}$). 
There is a slower fall off initially as $t_{1}$ starts deviating from $t$, however there is a near vertical decrease in the width as $t_{1}$ approaches $2t$, 
which eventually vanishes at the critical (semi-Dirac) point, namely, $t_{1} = 2t$.  

In order to visualize the dependence of $w$ on the band structure, we have depicted the conduction and the valence bands along the $k_x$-axis 
for $k_y = 2\pi/3$ for different values of $t_1$ in Fig. \ref{Chern_band}, where the $nnn$ hopping is absent, that is, we set $t_2 = 0$. 
The conduction and the valence bands are shown by the
red and the blue curves, while the values $\pm h_z/3\sqrt{3}t_2$ (see definition of $h_{z}$ in Eq. (\ref{kphamiltonian}) where the complex
$nnn$ hopping, $t_{2}$ enters through $h_{z}$ with $3\sqrt{3}t_{2}$ being the value of $h_{z}$ at the $\mathbf{K}^\prime$ and $\mathbf{K}^\prime$ points in the
Haldane model) are shown by the green and the purple 
curves.  The dotted vertical lines are drawn at the band touching points which depend on the $nn$ hopping strength $(t_1)$ and the height of 
these lines between the positive and the negative values of $h_z$ is $2w/w_H$, where $w$ is width of the Chern insulating phase for $t_1\neq t$. 
So the width depends on the magnitude of $h_z$ at the band touching point and  this magnitude decreases as one approaches the $\mathbf{M}$ point 
starting from the $\mathbf{K}$ and the $\mathbf{K}^\prime$ points. Mathematically, it can be represented as,
\begin{equation}
	\frac{w}{w_H} = \frac{h_z(k_x, k_y)}{3\sqrt{3}t_2}
\end{equation}
where $k_x$ and $k_y$ denote the magnitude of the momenta for which the energy gap vanishes. It should be noted that between 
$\mathbf{K}$ and $\mathbf{K}^\prime$, the band 
touching points for different $t_1$ can be evaluated by setting either $h_x$ or $h_y$ to zero and $k_y$ to $2\pi/3$ which results in the following equation,
\begin{equation}
k_x = \pm \frac{2}{\sqrt{3}}\cos^{-1}\left(\frac{t_1}{2t} \right),\hspace{15pt} 1 \leq t_1 \leq 2
\end{equation}
These boundary points are shown in Fig. \ref{Chern_band}.

\section{Anomalous Hall conductivity}

In this section we present numerical calculations of the Hall conductivity for our system. A non-zero local Berry curvature gives rise to the anomalous 
Hall conductivity. In order to calculate it we first obtain the low energy form for the tight binding Hamiltonian in Eq. (\ref{rawhamiltonian}) 
for different choices of $t_1$. Such a low energy expansion of the Hamiltonian will be valid and helpful for our purpose. It can be written in a compact 
notation as,
\begin{equation}\label{lowham}
\begin{aligned}
H = d_x (t_{1}/t,k_{x},k_{y})t\sigma_x &+ d_y(t_{1}/t,k_{x},k_{y})t\sigma_y \\&+ d_z(t_{1}/t,k_{x},k_{y})t_{2}\sigma_z
\end{aligned}
\end{equation}

\begin{center}
\begin{table}[h]
\begin{tabular}{|c||c|c|c|}
\hline
$t_1/t$ & $d_x$ & $d_y$& $d_z$\\ [0.5ex]
\hline\hline
1 & \parbox{1cm}{\begin{align*}
& \frac{3}{2} k_x-\frac{3}{8} k_x^2\\ &+ \frac{1}{4}k_y^2
\end{align*}} & \parbox{1cm}{\begin{align*}
\frac{3}{2} k_y-\frac{3}{8} k_y k_x
\end{align*}}
 & $-3\sqrt{3}$ \\[0.5ex]
\hline
1.9 & \parbox{1cm}{\begin{align*}
&\frac{3}{4} k_x^2 -\frac{7}{10}k_y^2 \\ & +\frac{29}{20}\sqrt{3}k_y -\frac{1}{20}
\end{align*}} & \parbox{1cm}{\begin{align*}
&\frac{9\sqrt{3}}{40}k_y^2 + \frac{29}{20} k_y\\&\frac{3\sqrt{3}}{4}k_x^2 + \frac{\sqrt{3}}{20}
\end{align*}} & $-4\sqrt{3}k_x$ \\[0.5ex]
\hline
2 & \parbox{1cm}{\begin{align*} \frac{3}{4}k_x^2 - \frac{3}{4} k_y^2
\end{align*}}& \parbox{1cm}{\begin{align*}
3k_y +& \frac{3\sqrt{3}}{4}k_x^2 \\ & - \frac{3}{4} k_y^2
\end{align*}}& $-4\sqrt{3}k_x$ \\[0.5ex]
\hline
2.1 & \parbox{1cm}{\begin{align*}
&\frac{3}{4} k_x^2 -\frac{8}{5}k_y^2 \\ & +\frac{31}{20}\sqrt{3}k_y +\frac{1}{20}
\end{align*}} & 
\parbox{1cm}{\begin{align*}
&\frac{11\sqrt{3}}{40}k_y^2 + \frac{31}{20} k_y\\&\frac{3\sqrt{3}}{4}k_x^2 - \frac{\sqrt{3}}{20}
\end{align*}} & $-4\sqrt{3}k_x$ \\[0.5ex]
\hline
\end{tabular}\caption{The coefficients $d_{x}$, $d_{y}$ and $d_{z}$ for a few different values of $t_1/t$ are shown in the table. The coefficients are presented up to terms quadratic in $k_x$ and $k_y$.}\label{tab1}
\end{table}
\end{center}
\begin{figure}[h]
	\centering
	\includegraphics[width = 0.42\textwidth]{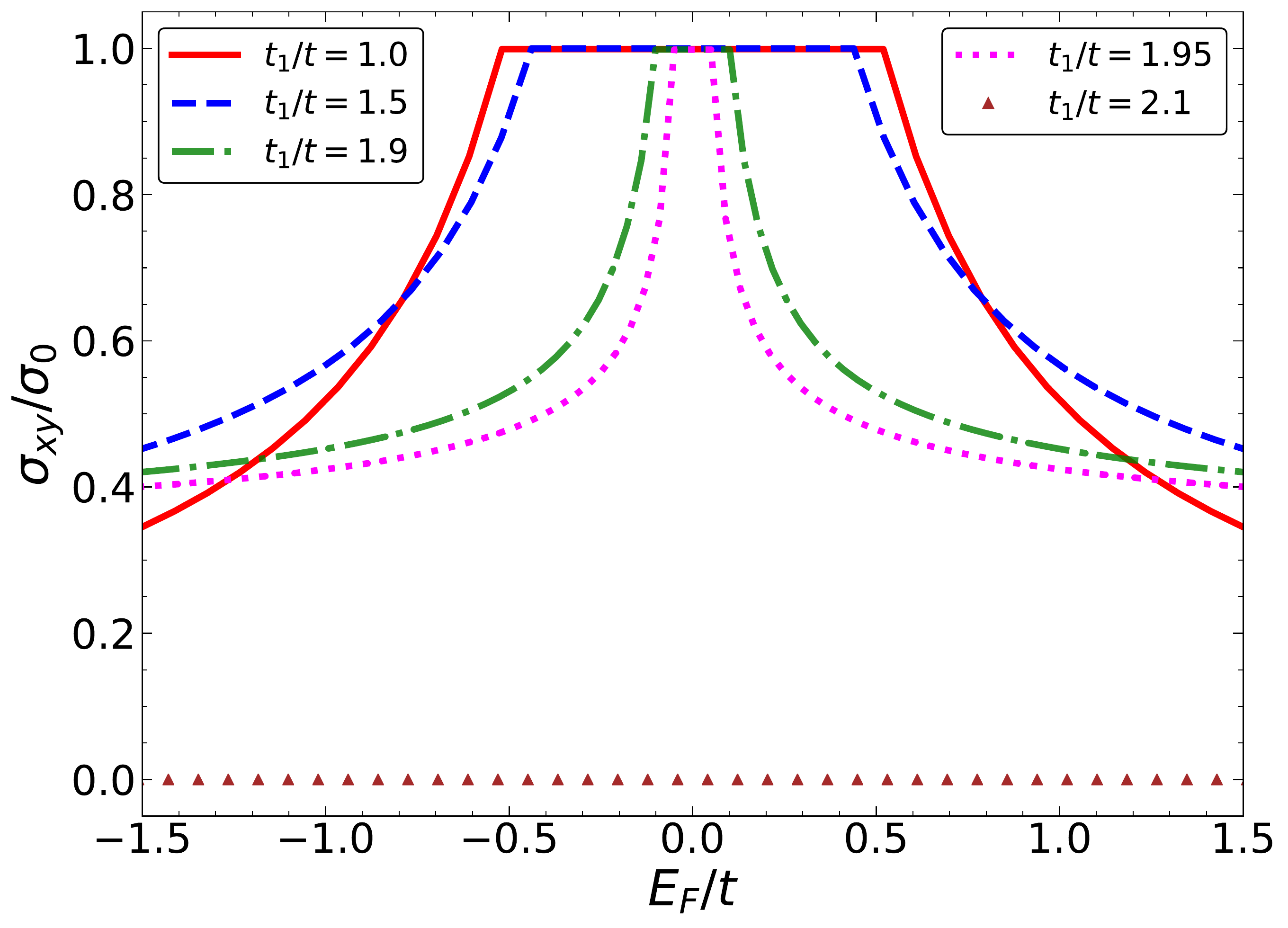}
	\caption{The variation of the Hall conductivity, $\sigma_{xy}$ is shown as a function of Fermi energy, $E_F$ for several values of $t_1$. 
Here $\sigma_0 = e^2/h$ is the unit of Hall conductance. In the calculation we have set $t_2=0.1t$ and $\phi=\pi/2$. It is evident from the figure that the 
plateau width decreases with increase in $t_1$ values.}
	\label{hallcond}
\end{figure}
where the coefficients $d_{x}$, $d_{y}$ and $d_{z}$ are functions of $k_{x}$, $k_{y}$ and $t_{1}/t$. These coefficients are presented up to terms quadratic in $k_x$ and $k_y$ corresponding to a few values of $t_{1}/t$ in table \ref{tab1}. For simplicity we have included the lattice constant in the definition of the momentum, \textbf{k} so that it is rendered dimensionless. It is 
to be kept in mind that the low energy expansions are done at different $k_{x}$, $k_{y}$ points for different values of $t_{1}$ since the bands come closest
at different points in the BZ. As an example, for $t_1 = t$, the approximation is done around the point with coordinates $(2\pi/3\sqrt{3}, 2\pi/3)$ 
(the Dirac points), whereas for $t_1 \le 2t$, the expansion is done around the point $(0, 2\pi/3)$ (the $\mathbf{M}$ point). 

The Hamiltonian with these coefficients in table \ref{tab1} represents the generic Hamiltonian for a two-level system and is of the form, 
$H = \mathbf{d}\cdot \boldsymbol{\sigma}$, where the components of $\mathbf{d}$ are the coefficients of $\sigma_i$s in Eq. (\ref{lowham}). 
One may distinguish the $d_{i}$s from the $h_{i}$s in Eq. (\ref{kphamiltonian}) where the latter, namely, $h_{x}$, $h_{y}$ and $h_{z}$ denote the coefficients corresponding to the full tight binding Hamiltonian. 
For $t_1 = t$, the coefficients of $\sigma_{x}$ and $\sigma_{y}$, namely, $d_x$ and $d_y$ depend on terms that are linear in $k_x$ and $k_y$ respectively.
For larger $t_{1}$, particularly for $t_1 = 2t$, $d_x$ is quadratic in both $k_x$ and $k_y$, while $d_y$ has terms linear in $k_y$ and quadratic in both $k_x$ and $k_y$. This renders non-uniformity in the band dispersion and hence
unequal velocities in different directions in the $k$-space. Further, the coefficient of $\sigma_z$, which arises because of the Haldane term (complex $nnn$ hopping)
is a constant for $t_1 = t$, while for larger values of $t_1$, it varies linearly in $k_x$. However for small $k$ values (near the $\mathbf{M}$ point), the linear terms are dominant which gives anisotropic linear band dispersion at low energies. The simple form of the Hamiltonian 
facilitates computation of the anomalous Hall conductivity using \cite{hall1, hall2},
\begin{equation}\label{eqhall}
\sigma_{xy} = \frac{e^2}{\hbar} \int \frac{d\mathbf{k}}{(2\pi)^2} f\left(E_{\mathbf{k}} \right) \Omega(\mathbf{k})
\end{equation}
where $\Omega(\mathbf{k})$ denotes the Berry curvature which can be obtained from the derivatives of the ${\textbf{d}}$ vector with respect to the
momenta, $k_{x}$ and $k_{y}$ using, \cite{kushsaha} 
\begin{equation}\label{berry}
	\Omega(\mathbf{k}) = \frac{\mathbf{d}}{2|\mathbf{d}|^3} \cdot \left( \frac{\partial \mathbf{d}}{\partial k_x} \times  \frac{\partial \mathbf{d}}{\partial k_y} \right).
\end{equation}
and $f(\epsilon ) = [e^{\beta(\epsilon -E_F)} +1]^{-1}$ is the Fermi-Dirac distribution function with $\beta$ being the inverse temperature ($\beta = 
1/k_{B}T$). 
\begin{figure}[h]
\centering
\includegraphics[width = 0.40\textwidth]{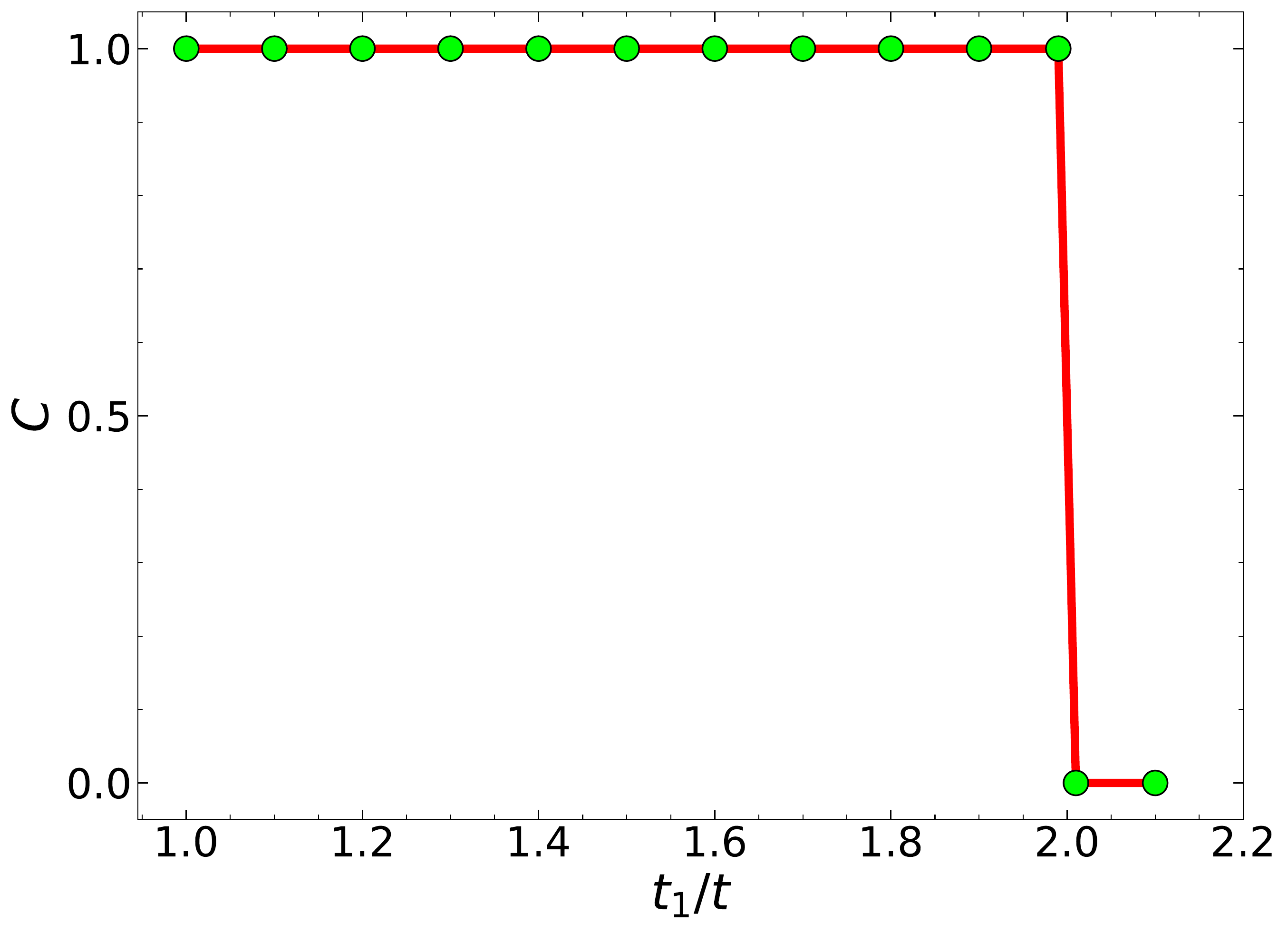}
\caption{The Chern number is plotted as a function of $t_1/t$. The Chern number stays at 1 for all values of $t_1<2t$ and vanishes for $t_{1} > 2t$.}
\label{Chernnoplot}
\end{figure}
\begin{figure}[h]
\centering
\includegraphics[width = 0.40\textwidth]{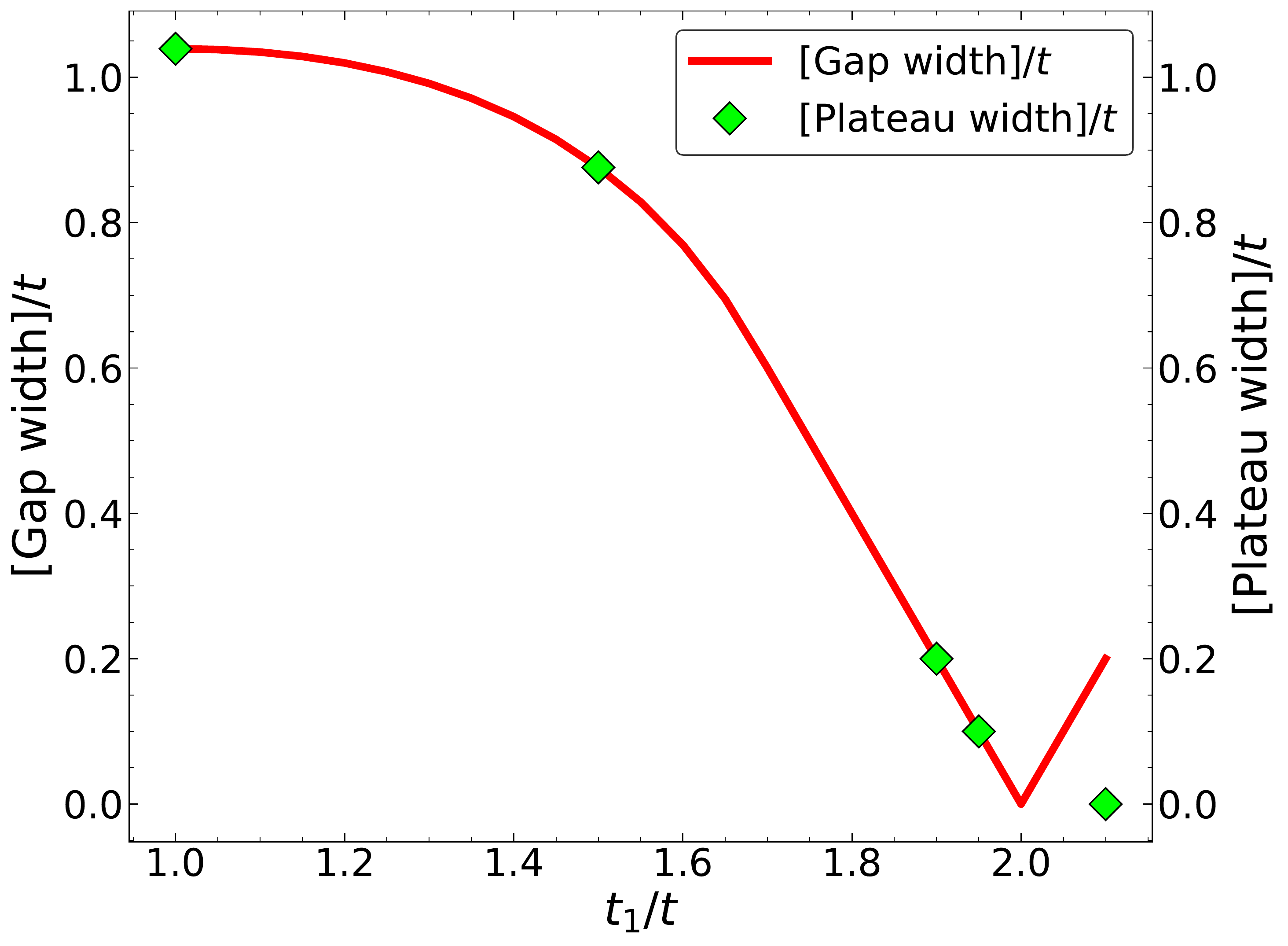}
\caption{The gap width and the plateau width in Hall conductivity are shown as a function of  $t_1/t$. They are labeled by the red dashed line and the 
green diamond points respectively. The plateau width is same as the gap width till $t_1 < 2t$. At $t_1 = 2.1t$ the width of the Hall plateau is 
zero but gap is non-zero.}
\label{gapwidth}
\end{figure}

The total Hall conductivity is the sum of $\sigma_{xy}$ from all the bands. Using equations (\ref{berry}) and (\ref{eqhall}) the Hall 
conductivity is calculated numerically and plotted as a function of the Fermi energy, $E_F$ for $T = 0$ in Fig. \ref{hallcond}. We can see that as long as the
the Fermi energy lies in the gap the Hall conductivity has a plateau and hence quantized in unit of $e^2/h$.
The integral in Eq. (\ref{eqhall}) is performed over the states which are partially occupied at 
a given value of the Fermi energy. Consequently, the Hall conductivity decreases rapidly as the Fermi energy moves away from the gapped region. 
In Fig. \ref{hallcond} the quantized plateau at $e^{2}/h$ stays, except that the plateau width decreases with increase in the value of $t_1$. 
This is because the width of the energy gap in the spectrum decreases with the increasing strength of $t_1$, although the Hall plateau survives as 
long as the Fermi energy lies in the gap. 

The above inference is supported by Fig. \ref{Chernnoplot} where we plot the Chern number, $C$ as a function of $t_{1}/t$. $C$ stays at 1 as long as
$t_{1}/t < 2$, implying continued existence of the topological phase, while as soon as $t_{1}/t$ becomes slightly larger than 2, $C$ drops to zero.
Thus we get a normal insulator beyond the semi-Dirac limit. 

We further derive support of the above scenario via Fig. \ref{gapwidth}, where we have shown the variation in the width of the Hall plateau and 
the width of the energy gap as a function of $t_1/t$. Here the band gap decreases with the increase in $t_1$ and the plateau width is 
found to be proportional to the energy gap. 
So the width of the Hall plateau follows the energy gap till $t_1 = 2t$. Again as $t_1$ becomes larger than $2t$, the plateau width vanishes, but the gap in the 
dispersion continues to exist.

\section{Conclusions}

We have investigated the evolution of the electronic spectrum and the topological properties in a Chern insulating model with tunable band dispersion. The tunable
dispersion is achieved by creating an anisotropy in hopping among a pair of nearest neighbours ($t_{1}$) as compared to the other two ($t$). The model shows a 
topological phase transition from a Chern insulating regime to a trivial insulating phase as the above anisotropy is progressively made larger. Computation 
of the electronic dispersion shows the Dirac points move towards each other and finally merge at the $\mathbf{M}$ point where the conduction 
and the valence bands touch each other as the hopping amplitudes satisfy $t_{1}=2t$. Eventually, for larger values of $t_{1}$, the 
spectrum gets gapped out again, where it shows properties of a trivial insulator. These conclusions are supported by computing the band dispersion, 
DOS, edge modes, Berry curvature, the Chern number phase diagram and the anomalous Hall conductivity. 

\begin{acknowledgments}

SB acknowledges helpful discussion with P. Kotetes.

\end{acknowledgments}


\begin{thebibliography}{8}
   \bibitem{Haldane1988}
	F. D. M. Haldane,
	\href{https://doi.org/10.1103/PhysRevLett.61.2015}{Phys. Rev. Lett. {\bf 61}, 2015 (1988).}
   \bibitem{Dietl2008}
   
   P. Dietl, F. Pichon, and G. Montambaux, 
   \href{https://doi.org/10.1103/PhysRevLett.100.236405} {Phys. Rev. Lett. {\bf 100}, 236405    (2008).}
   
   \bibitem{Banerjee2009}
   S. Banerjee, R. R. P. Singh, V. Pardo, and W. E. Pickett,
   \href{https://doi.org/10.1103/PhysRevLett.103.016402} {Phys. Rev. Lett. {\bf 103}, 016402  (2009).}
   
	\bibitem{pickett2009}
	V. Pardo and W.E. Pickett,
	\href{https://doi.org/10.1103/PhysRevLett.102.166803}{Phys. Rev. Lett. {\bf 102}, 166803 (2009)}
	
	\bibitem{pickett2010}
	V. Pardo, W.E. Pickett,
	\href{https://doi.org/10.1103/PhysRevB.81.035111}{Phys. Rev. B {\bf 81}, 035111 (2010)}
	
	\bibitem{castro_2014}
	A. S. Rodin, A. Carvalho, and A. H. Castro Neto,
	\href{https://doi.org/10.1103/PhysRevLett.112.176801}{Phys. Rev. Lett. {\bf 112}, 176801 (2014).}
	
	\bibitem{guan_2014}
	J. Guan, Z. Zhu, and D. Tománek,
	\href{https://doi.org/10.1103/PhysRevLett.113.046804}{Phys. Rev. Lett. {\bf 113}, 046804 (2014).}
	
	\bibitem{katnelson_2015}
	A. N. Rudenko, Shengjun Yuan, and M. I. Katsnelson,
	\href{https://doi.org/10.1103/PhysRevB.92.085419}{Phys. Rev. B {\bf 92}, 085419 (2015).}
	
	\bibitem{katnelson_2016}
	C. Dutreix, E. A. Stepanov, and M. I. Katsnelson,
	\href{https://doi.org/10.1103/PhysRevB.93.241404}{Phys. Rev. B {\bf 92}, 241404(R) (2016).}
	
	\bibitem{montambaux_2009}
	G. Montambaux, F. Piéchon, J.-N. Fuchs, and M. O. Goerbig,
	\href{https://doi.org/10.1103/PhysRevB.80.153412}{Phys. Rev. B {\bf 80}, 153412 (2009).}
	
	\bibitem{zhang_2017}
	C. Zhong, Y. Chen, Y. Xie, Y.-Y. Sun, and S. Zhang,
	\href{https://doi.org/10.1039/C7CP03558F}{Phys. Chem. Chem. Phys. {\bf {19}}, 3820 (2017).}
	
	\bibitem{Suzumura2013}
	Y. Suzumura, T. Morinari, and F. Pichon,
	\href{https://doi.org/10.7566/JPSJ.89.023701}{ J. Phys. Soc. Jpn. {\bf 82}, 023708 (2013).}
	
	\bibitem{kohmoto_2006}
	Y. Hasegawa, R. Konno, H. Nakano, and M. Kohmoto,
	\href{https://doi.org/10.1103/PhysRevB.74.033413}{Phys. Rev. B {\bf 74}, 033413 (2006).}
	
	\bibitem{bryenton2019}
	J. P. Carbotte, K. R. Bryenton, and E. J. Nicol
	\href{https://doi.org/10.1103/PhysRevB.99.115406}{Phys. Rev. B {\bf 99}, 115406 (2019).}

	\bibitem{mawrie2019}
	A. Mawrie and B. Muralidharan,
	\href{https://doi.org/10.1103/PhysRevB.99.075415}{Phys. Rev. B {\bf 99}, 075415 (2019).}
	
	\bibitem{Narayan2015}
	A. Narayan,
	\href{https://doi.org/10.1103/PhysRevB.91.205445}{Phys. Rev. B {\bf 91}, 205445 (2015).}
	
	\bibitem{Chen2018}
	Q. Chen, L. Du, and G. Fiete, 
	\href{https://doi.org/10.1103/PhysRevB.97.035422}{Phys. Rev. B {\bf 97}, 035422 (2018).}
	
    \bibitem{Firoz2018}
    S. K. F. Islam and A. Saha,	
    \href{https://doi.org/10.1103/PhysRevB.98.235424}{Phys. Rev. B {\bf 98}, 235424 (2018).}
    
    
    \bibitem{Mandal2020}	
    I. Mandal and K. Saha, 
    \href{https://doi.org/10.1103/PhysRevB.101.045101}{Phys. Rev. B {\bf 101}, 045101 (2020).}
    
    \bibitem{Mawrie2019(2)}
    A. Mawrie and B. Muralidharan
    \href{https://doi.org/10.1103/PhysRevB.100.081403}{Phys. Rev. B {\bf 100}, 081403 (2019).}
    
	
	\bibitem{banerjee2012}
	S. Banerjee and W.E. Pickett,
	\href{https://doi.org/10.1103/PhysRevB.86.075124}{Phys. Rev. B {\bf 86}, 075124 (2012).}
  	 
   \bibitem{Zhou2015}
    X. Y. Zhou, R. Zhang, J. P. Sun, Y. L. Zou, D. Zhang, W.
    K. Lou, F. Cheng, G. H. Zhou, F. Zhai, and K. Chang,
    \href{https://doi.org/10.1038/srep12295}{Sci. Rep. {\bf 5}, 12295 (2015).}
    
    \bibitem{Yuan2016}
   S. Yuan, E. v. Veen, M. I. Katsnelson, and R. Roldan,
    \href{https://doi.org/10.1103/PhysRevB.93.245433}{Phys. Rev. B {\bf 93}, 245433 (2016).}
    
	\bibitem{sinha2020}
	P. Sinha, S. Murakami and S. Basu,
	\href{https://doi.org/10.1103/PhysRevB.102.085416}{Phys. Rev. B {\bf 102}, 085416 (2020)}

	\bibitem{semenhoff}
	G. Semenoff,
	\href{https://doi.org/10.1103/PhysRevLett.53.2449}{Phys. Rev. Lett. {\bf 53}, 2449 (1984).}
	
	\bibitem{Vanderbilt}
	T. Thonhauser and D. Vanderbilt,
	\href{https://doi.org/10.1103/PhysRevB.74.235111}{Phys. Rev. B {\bf 74}, 235111 (2006).}
	
	\bibitem{nakada1996}
	K. Nakada, M. Fujita, G. Dresselhaus, and M. S. Dresselhaus,
	\href{https://doi.org/10.1103/PhysRevB.54.17954}{Phys. Rev. B {\bf 54}, 17954 (1996).}
	
	\bibitem{castroneto}
	A. H. Castro Neto, F. Guinea, N. M. R. Peres, K. S. Novoselov, and A. K. Geim,
	\href{https://doi.org/10.1103/RevModPhys.81.109}{Rev. Mod. Phys. {\bf 81}, 109 (2009).}

	\bibitem{basu2018}
	P. Sinha, S. Ganguly and S. Basu,
	\href{https://doi.org/10.1016/j.physe.2018.06.005}{Physica E {\bf 103}, 314 (2018).}
	
	\bibitem{Kapri2020}
	B. Dey, P. Kapri, O. Pal, and T. K. Ghosh,
	\href{https://doi.org/10.1103/PhysRevB.101.235406}{Phys. Rev. B {\bf 101}, 235406 (2020).}
	
	\bibitem{Fukui}
	T. Fukui, Y. Hatsugai and H. Suzuki,
	\href{https://journals.jps.jp/doi/10.1143/JPSJ.74.1674}{J. Phys. Soc. Jpn. {\bf 74}, 1674 (2005).}
	
	\bibitem{thouless}
	D. J. Thouless, Topological Quantum Numbers in Nonrelativistic
	Physics \href{https://books.google.co.in/books?id=BgbtCgAAQBAJ}{(World Scientific, Singapore, 1998)}.
	
	\bibitem{hall1}
	D. Xiao, M.-C. Chang, and Q. Niu,
	\href{https://doi.org/10.1103/RevModPhys.82.1959}{Rev. Mod. Phys. {\bf 82}, 1959 (2010).}
	
	\bibitem{hall2}
	D. Culcer, A.MacDonald, and Q. Niu,
	\href{https://doi.org/10.1103/PhysRevB.68.045327}{Phys. Rev. B {\bf 68}, 045327 (2003).}
	
	\bibitem{kushsaha}
	K. Saha,
	\href{https://doi.org/10.1103/PhysRevB.94.081103}{Phys. Rev. B {\bf 94}, 081103(R) (2016).}
	
\end{thebibliography}
\end{document}